\documentclass[%
 reprint,
 amsmath,amssymb,
 aps,
]{revtex4-2}

\usepackage[dvipdfmx]{graphicx,color,hyperref}
\hypersetup{
	colorlinks=true,
	citecolor=blue,
	urlcolor=blue,
}
\usepackage{dcolumn}
\usepackage{bm}
\usepackage{blkarray}

\begin{document}
\preprint{APS/123-QED}

\title{A multicritical point of $S=1$ XXZ chain with  single ion anisotropy}

\author{Shuichi Shiraishi}
 \email{s.shiraishi@stat.phys.kyushu-u.ac.jp}
\author{Kiyohide Nomura}%
 \email{knomura@stat.phys.kyushu-u.ac.jp}
\affiliation{Department of Physics, Kyushu University,Fukuoka 819-0395,Japan \\
}%

\date{\today}

\begin{abstract} 
We research the ground state of the $S = 1$ XXZ spin chain with single ion anisotropy, focusing on XY1, XY2 (Spin Nematic), N\'eel, Haldane phases. Recently, it was reported that the four phases did not intersect at a single point using conventional numerical methods, which is inconsistent with the bosonization theory. In order to resolve this discrepancy, we propose a unified method to determine phase boundaries by combining conformal field theory (CFT) with numerical diagonalization. And we discuss the symmetry and universality class of the multicritical point.\\

\end{abstract}

\maketitle

\section{Introduction}
$S = 1$ XXZ spin chain with single ion anisotropy, which is given by

\begin{equation}\label{hamiltonian}
H=\sum_{j}({S}_{j}^{x}{S}_{j+1}^{x}+{S}_{j}^{y}{S}_{j+1}^{y}+{\Delta}{S}_{j}^{z}{S}_{j+1}^{z}+D({S}_{j}^{z})^{2})
\end{equation}
has been studied by many authors and especially, in detail by Schulz \cite{Schulz-1986} and Chen, Hida, Sanctuary \cite{Chen-Hida-Sanctuary-2003}.
According to their studies, the model has several different phases in the ground state depending on the Hamiltonian parameters. In particular, there are regions where four phases, called the XY1, XY2, Haldane, and N\'eel phases, are adjacent to each other.\par

 At first we overview phases of model (\ref{hamiltonian}). Ferromagnetic phase is characterized with the ferromagnetic long-range order,
and N\'eel phase with the N\'eel long-range order. On the other hand, there is no long-range order in the Haldane, large D, XY1 and XY2 phases.
The correlation lengths are finite in the Haldane and the large D phases,
though correlation lengths diverge in the XY1 and XY2 phases (or massless phases).
Universality class of 
the XY1-Haldane and the XY1-large D phase transition
belongs to the Berezinskii-Kosterlitz-Thouless (BKT) transition type,
and the Haldane-large D phase transition belongs to the Gaussian fixed line associated with BKT transition.\par

In Schulz's work, analytical methods applying bosonization were used, while the work of Chen et al. was well investigated numerically.
However, Chen et al, used two different methods for N\'eel-Haldane phase boundary and for XY1-XY2. This is because the method for an order-disorder transition is inappropriate for determining the XY1-XY2 phase boundary. In this case, recent numerical study \cite{Tonegawa-Okamoto-Nomura-Sakai-2022} suggested that the four phases did not intersect at a single point, a result that is inconsistent with Schulz's work.
 By the way, since the model (\ref{hamiltonian}) has two parameters  $\Delta$ and $D$, the phase boundary between two phases is a line, the phase boundary between three phases is a point.
But for four phases, there is not intersection in general. Thus, we should check the Schulz's statement.\par

 The main purpose of this study is to investigate the physics around this multiple critical point using a unified method, and we discuss the critical phenomena by obtaining the energy with numerical diagonalization and conformal field theory.\par
\begin{figure}[hbtp]
 \centering
 \includegraphics[keepaspectratio, scale=0.45]
      {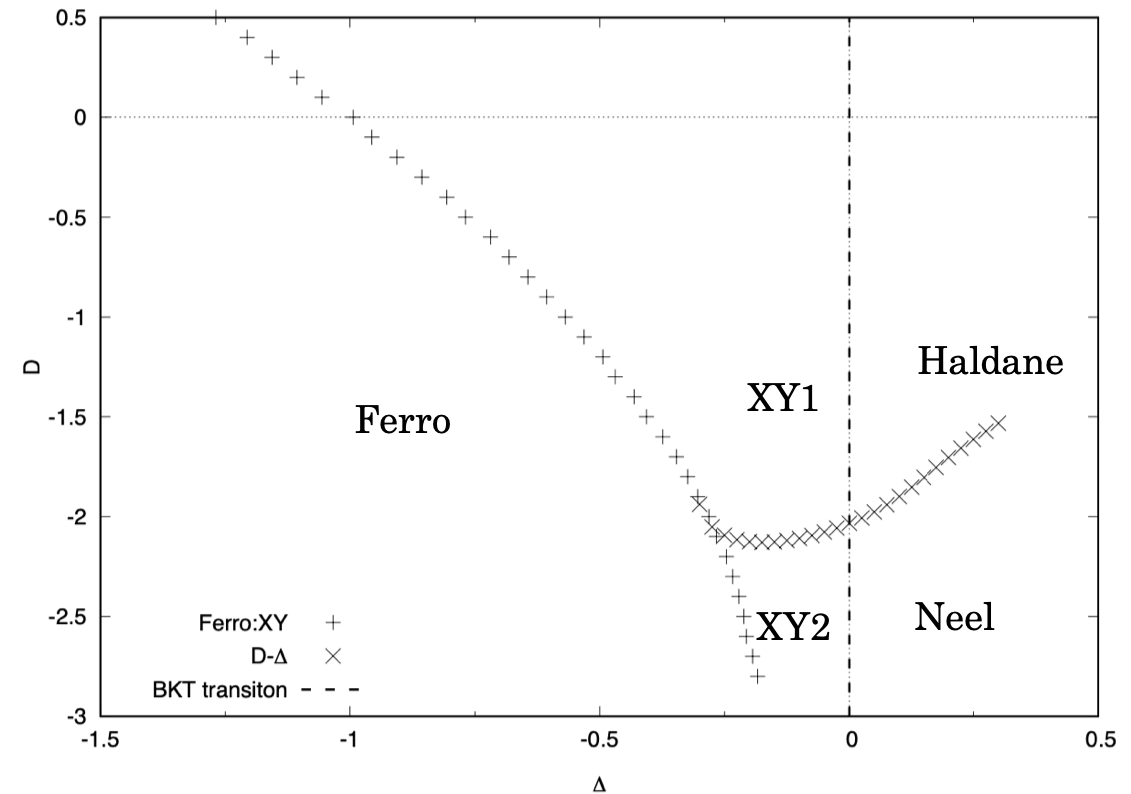}
 \caption{Phase diagram : Haldane-N\'eel and the XY1- XY2 phase boundaries are connected continuously and four phases intersect as the tetracritical point.}
\label{phase}
\end{figure}

 Using the method described in Section 3, we obtain a phase diagram in FIG. \ref{phase}. Here the XY1-Haldane and the XY2-N\'eel boundaries are on the $\Delta=0$ line. The Haldane-N\'eel and XY1-XY2 phase boundaries are connected smoothly, and four phases intersect at the tetracritical point $\Delta=0$, $D=-2.0351\pm0.0034$.

  Note that the line of $\Delta=0$ is BKT transition. This is supported by using the special SU(2) symmetry argument in the work of Kitazawa, Hijii, and Nomura \cite{Kitazawa-Hijii-Nomura-2003}. According to their study, in addition to the SU(2) symmetry of the conventional spin operators, there exists a special SU(2) symmetry. At first it is shown for $S = 1$
  \begin{equation*}\label{s-s1}
 \lbrack S_{j}^{z} ,(S_{k}^{\pm})^{2} \rbrack = \pm 2\delta_{jk} (S_{j}^{\pm})^{2}
,\,\,\,
\lbrack (S_{j}^{+})^{2} ,(S_{k}^{-})^{2} \rbrack = 4 \delta_{jk} S_{j}^{z}.
\end{equation*}
Then, they introduced 
\begin{equation}\label{s2}
s_{j}^{\pm}=\frac{1}{2}(S^{\pm}_{j})^{2}U_{j},\,\,\,
s^{z}_{j}=\frac{1}{2}S^{z}_{j}
\end{equation}
where $U_{j}$ is the non-local unitary operator
\begin{equation*}\label{U_{j}}
U_{1}=1,\,\,\,
U_{j}=\prod_{l=1}^{j-1}(1-2(S^{z}_{l})^{2}) =e^{i\pi \sum_{l=1}^{j-1}S^{z}_{j}}.
\end{equation*}
From the commutation relation $[(S^{z}_{j})^{2},(S^{\pm}_{j})^{2}]=0$, they obtained the relations as below
\begin{equation}\label{s4}
[s^{z}_{j},s^{\pm}_{k}]=\pm \delta_{jk}s^{\pm}_{j},\,\,\,
[s^{+}_{j},s^{-}_{k}]=2\delta_{jk}s^{z}_{j}.
\end{equation}
When $\Delta=0$, the Hamiltonian (\ref{hamiltonian}) and $s^{\pm}_{T}=\sum_{j=1}^{L}s^{\pm}_{j}$  commute.\par
One of the authors proposed the level spectroscopy technique \cite{Nomura-1995,Nomura-Kitazawa-1998} to treat BKT transition,
by combining renormalization group, symmetry and numerical data. 
For the Gaussian fixed line, Kitazawa proposed the twist boundary method \cite{Kitazawa-1997}.
However, the distinction between the XY1 and XY2 phases is subtle,
which we will explain in the following sections.\par

  \par

  \par

We review a short history of the XY1 and XY2 phases.
By using perturbation and numerical diagonalization,
Solyom and Ziman \cite{Solyom-Ziman-1984}
studied S=1 XXZ spin chain with single ion anisotropy and discussed the planar phase, but they did not distinguish the two (XY1-XY2) phases.\par
Later,
combining numerical diagonalization and bosonization, 
Schulz and Ziman \cite{Schulz-Ziman-1986}
discussed there are two phases in massless phases;
planar1 (XY1) and planar2 (XY2).
Schulz \cite{Schulz-1986}
explained the bosonization technique for the XXZ chain with single ion anisotropy for general spin S cases, where he expressed spin S chain by the spin ladder with ferromagnetic rung couplings, and then studied two continuous effective field models. One effective field model induces the BKT transition, and another induces the 2D Ising type transition, thus there appear four phases.\par
Chen et al.\cite{Chen-Hida-Sanctuary-2003}
studied the S=1 XXZ spin chain with single ion anisotropy, using numerical diagonalization
and the level spectroscopy technique \cite{Nomura-Kitazawa-1998}, 
and they found that the BKT transition line
(XY1-Haldane and XY2-N\'eel)
is located just on the $\Delta=0$.
They also determined the large D-Haldane phase boundary,
by using the twist boundary method.\par
However, since they used the different methods to determine the Haldane-N\'eel phase boundary (phenomenological renormalization group method; PRG\cite{Fisher and Barber})
and the XY1-XY2 phase boundary, thus these two phase boundaries may not connect continuously for finite system. In fact, recently, Tonegawa et al.\cite{Tonegawa-Okamoto-Nomura-Sakai-2022}
studied S=1/2 ferromagnetic-antiferromagnetic bond-alternating chain, which is considered to belong to the same universality type as S=1 XXZ chain with single ion anisotropy, by analyzing numerical diagonalization. Especially in the four phases, that is, spin nematic Tomonaga-Luttinger (nTLL), up-up-down-down (UUDD), singlet-dimer (SD) phases, where nTLL phase corresponds to XY2 phase, UUDD to N\'eel, SD to Haldane, they found several tricritical points.
We consider that such tricritical points may be artificial one coming from
different numerical methods,
thus we propose a new unified method to determine the XY1-XY2 phase boundary and 
the Haldane-N\'eel phase boundary.\par

Let us mention one model  which shows such a tetracritical behavior;
the bilinear-biquadratic (BLBQ) model with single ion anisotropy
\begin{equation}\label{eq:BLBQ-w-sigle-ion-anisotropy}
  H= \sum_{j} \left(\cos \theta (\bm{S}_{j}\cdot \bm{S}_{j+1})+\sin \theta (\bm{S}_{j}\cdot \bm{S}_{j+1})^{2}\right)
  + D \sum_{j} (S^{z}_{j})^{2}.
\end{equation}
At the Takhtajan-Babujian (TB) point $D=0$ and $\theta=- \pi/4$, this model is Bethe-Ansatz solvable \cite{Takhtajan-1982,Babujian-1982,Babujian-1983}.
Field theoretical models at this point were proposed by
\cite{Affleck-Haldane-1987} and
\cite{Tsvelick-1990}.
Numerically, it was found that the four phases (Haldane, dimer, N\'eel, large D) 
meet at the TB point
\cite{DeChiara-Lewenstein-Sanpera-2011,Lepri-DeChiara-Sanpera-2013}
.
Then, the following question arises: whether the universality class of the tetracritical point of model (\ref{hamiltonian}) is the same with that of the TB point or not?

In Sec.II, we summarize Schulz's discussion. In Sec.III, we carry out numerical calculations, and propose a new method. In Sec.IV, we discuss symmetry aspects of this model (\ref{hamiltonian}). In Sec.V, the conclusions of this paper are shown. In the Appendix, we overview the effective Hamiltonian based on the perturbation calculations.

\section{Review of Schulz's discussion}

Schulz pointed out in reference \cite{Schulz-1986} that there are two different types of XY phases with different characteristics. First, the XY1 and XY2 phases are characterized by the following spin correlation functions, which show both a power-law decay and an exponential decay respectively.
In XY1 phase, they are
 \begin{equation}\label{xy1}
\begin{split}
 \langle S_{j}^{+} S_{j+r}^{-} \rangle &\propto (-1)^{r} r^{-{\eta}_{1}}\\
 \langle (S_{j}^{+} S_{j}^{+}) (S_{j+r}^{-}S_{j+r}^{-}) \rangle &\propto r^{-{\eta}_{2}}\\
 \langle S_{j}^{z} S_{j+r}^{z} \rangle &= C_{z} r^{-2} + D_{z} (-1)^{r} \exp(-r/{\xi_{z}})
  \end{split}
\end{equation} \\
where, $0 < \eta_{1} < 1/4$ and $\eta_{2}=4\eta_{1}$. Then, on the boundary of XY1-Haldane, they are $\eta_{1}= 1/4$, $\eta_{2}=1$.\par
In XY2 phase,
\begin{equation}\label{xy2}
\begin{split}
 \langle S_{j}^{+} S_{j+r}^{-} \rangle &\propto (-1)^{r} \exp(-r/{\xi_{1}})\\
 \langle (S_{j}^{+} S_{j}^{+}) (S_{j+r}^{-}S_{j+r}^{-}) \rangle &\propto r^{-{\eta}_{2}}\\
 \langle S_{j}^{z} S_{j+r}^{z} \rangle &\propto (-1)^{r} r^{-{\eta}_{z}}
  \end{split}
\end{equation}\\
where $0<\eta_{2}<1$, $\eta_{z}=1/\eta_{2}$, and on the boundary of XY2-N\'eel, $\eta_{2}= 1$.

\par
Furthermore, correlation functions are in Haldane phase,
\begin{equation}\label{haldane}
\begin{split}
 \langle S_{j}^{+} S_{j+r}^{-} \rangle &\propto (-1)^{r} \exp(-r/{\xi_{1}})\\
 \langle (S_{j}^{+} S_{j}^{+}) (S_{j+r}^{-}S_{j+r}^{-}) \rangle &\propto \exp(-r/{\xi_{2}})\\
 \langle S_{j}^{z} S_{j+r}^{z} \rangle &\propto (-1)^{r} \exp(-r/{\xi}_{z}),\\
  \end{split}
\end{equation}\\
whereas in N\'eel phase,
\begin{equation}\label{neel}
\begin{split}
 \langle S_{j}^{+} S_{j+r}^{-} \rangle &\propto (-1)^{r} \exp(-r/{\xi_{1}})\\
 \langle (S_{j}^{+} S_{j}^{+}) (S_{j+r}^{-}S_{j+r}^{-}) \rangle &\propto \exp(-r/{\xi_{2}})\\
 \langle S_{j}^{z} S_{j+r}^{z} \rangle &\propto (-1)^{r}.\\
  \end{split}
\end{equation}\\

From the lattice model (\ref{hamiltonian}), Schulz deduced two independent field models. One field model induces the BKT transition, whereas another induces the 2D Ising type transition. Therefore he concluded the four phases (XY1, XY2, Haldane, N\'eel) intersect at one point.\par

By the way, in the conformal field theory (CFT) language, the XY1 and the XY2 phases belong to the central charge $c=1$. The Haldane-N\'eel phase boundary belongs to the $c=1/2 $ CFT (or 2D Ising type universality). For a review of conformal field theory, reference \cite{Ginsparg} is helpful.

Finally, with perturbative calculations  in  $D \rightarrow -\infty$ limit as described  in appendix,
one can obtain an effective model which has  the ferromagnetic, XY2 and N\'eel phases.
This is consistent with  the above mentioned Schulz's discussion.

\section{Numerical analysis}
In this section, we show the numerical analysis. The energy of system is calculated by numerical diagonalization. Considering dispersion relation, we define wave number as
\begin{equation}\label{q}
q=\frac{2 \pi}{L}n,
\end{equation}
and spin wave velocity as
\begin{equation}\label{v0}
v_{0}= \left. \frac{dE(q)}{dq} \right|_{q=0},
\end{equation}
magnetization  as the sum of the $z$ component spins,
\begin{equation}
m=\sum_{j}S_{j}^{z}
\end{equation}
 as follows. Where, $n$ is integer and $E(q)$ is the  energy at magnetization $m=0$ and wavenumber $q$. 
And also, we define the translation operator $\hat{T}$ as acting like
\begin{equation}\label{t1}
\hat{T}^{\dag} \hat{S_{j}} \hat{T}= \hat{S}_{j+1}
\end{equation}
so that under periodic boundary conditions, 
\begin{equation}\label{t2}
\begin{split}
\hat{T}^{L} |phys \rangle &= \hat{T}^{L-1} \hat{T} |phys \rangle \\
&=\hat{T}^{L-1}e^{iq}|phys \rangle \\
&=e^{iq}\hat{T}^{L-1}|phys \rangle \\
&=e^{iqL} |phys \rangle \\
&=|phys \rangle
\end{split}
\end{equation}
where $L$ is the number of sites and $|phys \rangle$ denotes the eigenstates of $\hat{T}$.
\par
Now, since the spin wave velocity is defined by the derivative, ground state energy $E_{g}$ and $v_{0}$ can be specifically calculated numerically by replacing and treating it as the difference \begin{equation}\label{v0l}
v_{0}(L)=\frac{E(q = 2 \pi/L)-E(q = 0)}{2 \pi /L}
\end{equation}

\subsection{Effective central charge}
Previous numerical studies have shown that the central charge $c=1$ in XY1 \cite{Kitazawa-Nomura-1997} and $c=1$ in XY2 phase \cite{Chen-Hida-Sanctuary-2003}(see also appendix). In this subsection, we would like to confirm this fact and investigate the central charge at the phase boundaries. Here we take the approach of confirming the singularity by analyzing how the central charge $c$ depends on $D$, which is an anisotropic parameter. First, it is known that when conformal field theory holds, 

\begin{equation}\label{energy1}
\frac{E_{g}(L)}{L}={\epsilon}_{\infty}-\frac{{\pi}v_{0}c}{6L^{2}}, \ \ \ \frac{E_{g}(L-2)}{L-2}={\epsilon}_{\infty}-\frac{{\pi}v_{0}c}{6(L-2)^{2}}
\end{equation}
from the discussion on finite size scaling, \cite{Blote-1986} \cite{Affleck-1986} where $E_{g}$ is the ground state energy, $L$ is the system size, and $\epsilon_{\infty}$ is the ground energy per unit site in the thermodynamic limit.\par
Considering the difference in equation (\ref{energy1}), we can get the relation

\begin{equation}\label{c}
\begin{split}
c(L,L-2)&=\left\{\frac{E_{g}(L)}{L}-\frac{E_{g}(L-2)}{L-2}\right\}\\
&\times \left\{ -\frac{\pi}{6}\left(\frac{v_{0}(L)}{L^{2}}-\frac{v_{0}(L-2)}{(L-2)^{2}}\right) \right\}^{-1}.
\end{split}
\end{equation}
Thus, $c(L,L-2)$ can be obtained by numerical data $E_{g}(L)$ and $v_{0}(L)$, and we shall call it as the effective central charge.

\begin{figure}[hbtp]
 \centering
 \includegraphics[keepaspectratio, scale=0.32]
      {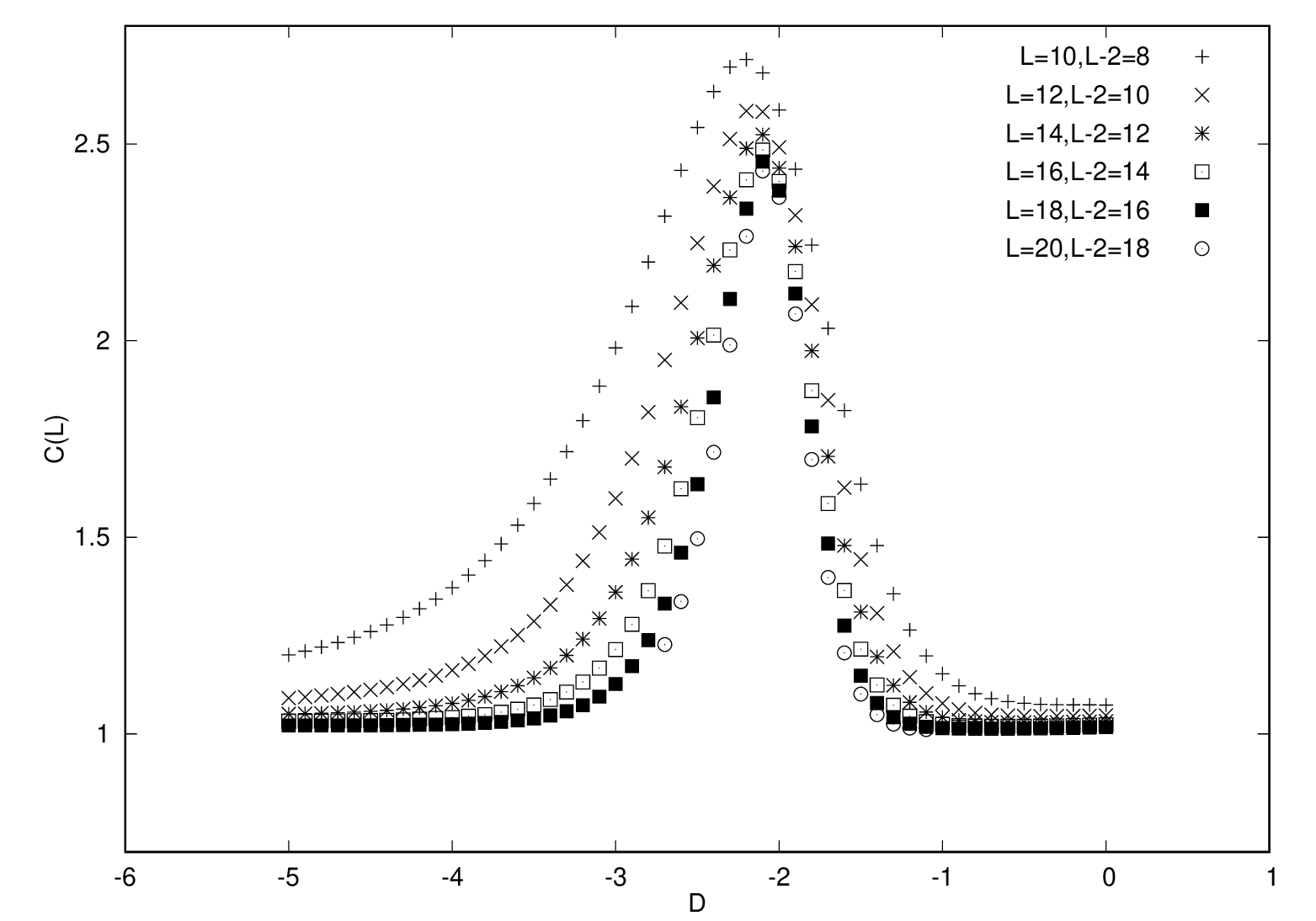}
\caption{Effective central charge at $\Delta =0$: In the region $D \neq -2$, c takes the predicted value, but at the region around $D \sim -2$ it differs from the naive expectation.}
\label{c-d}
\end{figure}

\begin{figure}[hbtp]
 \centering
 \includegraphics[keepaspectratio, scale=0.65]
      {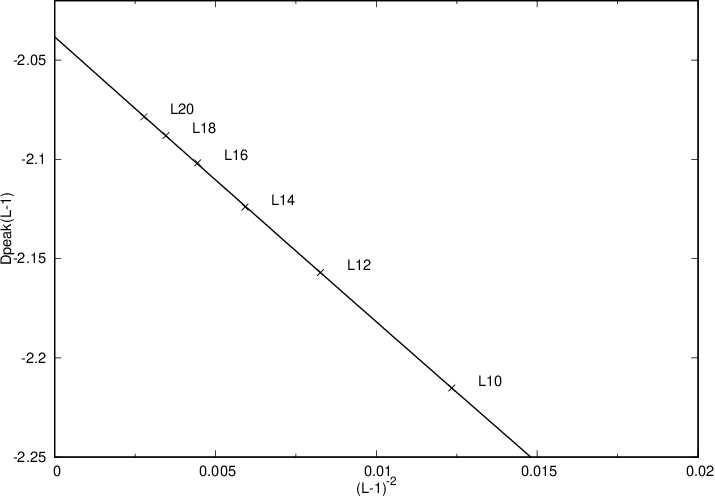}
 \caption{The peak of $D$ identification by central charge at $\Delta = 0$.}
\label{DpbyC}
\end{figure}

\begin{figure}[hbtp]
 \centering
 \includegraphics[keepaspectratio, scale=0.65]
      {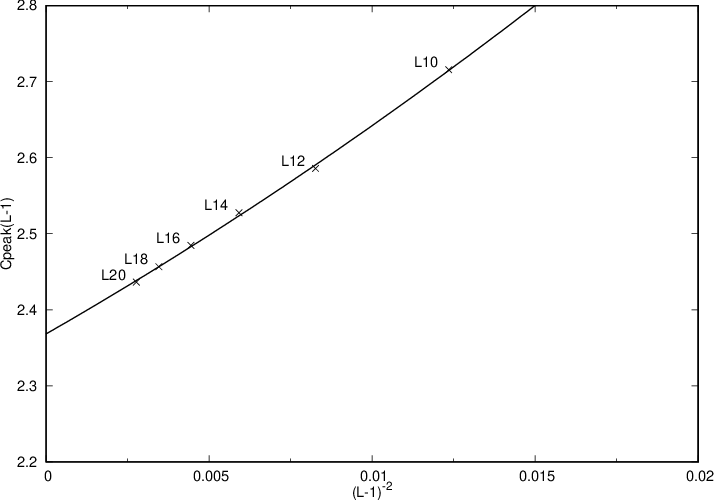}
\caption{The extrapolation of the peak of effective central charge at $\Delta = 0$.}
\label{charge}
\end{figure}

 FIG. \ref{c-d} plots the curves for each system size $L$, with $D$ on the horizontal axis and the central charge $c(L)$ on the vertical axis.
The results of numerical calculations using the exact diagonalization method are used for the bottom energies for system sizes $L = $ 8, 10, 12,14,16 and 18.
 The results of extrapolation to obtain the value of $D$ when $c$ takes a peak and the value of $c$ at that time are shown in FIG. \ref{DpbyC} and FIG. \ref{charge}. Here we extrapolated the effective central charge as 
\begin{equation}
c_{peak}(L-1) = c_{peak}(\infty) + \alpha_{1} \frac{1}{(L-1)^{2}}+ \alpha_{2} \frac{1}{(L-1)^{4}},
\end{equation}
and also extrapolated $D$ as $D_{peak}(L-1) = D_{peak}(\infty) + \beta_{1} \frac{1}{(L-1)^{2}}+ \beta_{2} \frac{1}{(L-1)^{4}}$ \cite{Cardy-1986},\cite{Reinicke-1986}.
 \par

From Fig. \ref{c-d}, we read the effective central charge becomes $c=1$,
except for $D\sim -2$. In Fig. \ref{DpbyC}, we plot the peak positions of the effective central charge as a function of size,
and we extrapolate the tetracritical point $D=-2.03686, \Delta=0$ in the infinite limit.
  Assuming that the central charge of one phase is $c_{1}$, the other is $c_{2}$, and the spin wave velocity $v_{0}$ is common at the phase boundary, the following simple prediction can be made,
 
 \begin{equation}\label{energy2}
\frac{E_{g}(L)}{L}={\epsilon}_{\infty}-\frac{{\pi}v_{0}(c_{1}+c_{2})}{6L^{2}}.
\end{equation}
On the basis of this relationship, we may expect the central charge $c=3/2$
at the tetracritical point, which is the sum of $c=1$ and $c=1/2$. 
For the TB point of (\ref{eq:BLBQ-w-sigle-ion-anisotropy}), this assertion is correct,
since the $c=1/2$ line (on  the Haldane-N\'eel and dimer-large D phase boundaries) 
and the $c=1$ line (on  the Haldane-large D and N\'eel-dimer phase boundaries)
cross at the TB point with central charge $c=3/2$
\cite{Affleck-Haldane-1987,Tsvelick-1990}.

In contrast, at the tetracritical point of the present model (\ref{hamiltonian}),
the effective central charge $c = 2.368$ obtained in Fig. 2 and 4,
deviates largely from $c=3/2$.
We consider that this discrepancy comes from the assumption of a unique spin wave velocity.

\subsection{Massless case}
Even in the case where multiple spin-wave velocities appear, the problem in the previous section can be avoided by using relation (\ref{energy3-1}) at three system sizes, $L+2,L,L-2$. In other words, we consider estimating the phase boundaries by the ratios as explained below, without directly obtaining the central charge.

\subsubsection{Definition of the 'ratio'}
If there exist two spin wave velocities $v_{1}$ and $v_{2}$, 
\begin{equation}\label{energy3-1}
 \frac{E_{g}(L)}{L}={\epsilon}_{\infty}-\frac{{\pi}}{6L^{2}}(v_{1}c_{1}+v_{2}c_{2}) ,
\end{equation}
would be viable instead of (\ref{energy1}). This relation may be written as
 \begin{equation}\label{energy3-2}
\frac{E_{g}(L+2)}{L+2}-\frac{E_{g}(L)}{L}=-\frac{\pi}{6}\left(\frac{1}{(L+2)^{2}}-\frac{1}{L^{2}}\right)(v_{1}c_{1}+v_{2}c_{2})
\end{equation}
and
\begin{equation}\label{energy3-3}
\frac{E_{g}(L)}{L}-\frac{E_{g}(L-2)}{L-2}=-\frac{\pi}{6}\left(\frac{1}{L^{2}}-\frac{1}{(L-2)^{2}}\right)(v_{1}c_{1}+v_{2}c_{2}).
\end{equation}
From (\ref{energy3-2}),(\ref{energy3-3}) , we obtain the relation
\begin{equation}\label{energy3-4}
\begin{split}
\left( \left.\cfrac{\cfrac{E_{g}(L+2)}{L+2}-\cfrac{E_{g}(L)}{L}}{\cfrac{E_{g}(L)}{L}-\cfrac{E_{g}(L-2)}{L-2}} \right)
\right/ \left( \cfrac{\cfrac{1}{(L+2)^{2}}-\cfrac{1}{L^{2}}}{\cfrac{1}{L^{2}}-\cfrac{1}{(L-2)^{2}}} \right)
=1.
\end{split}
\end{equation}
The relation (\ref{energy3-4}) is established where conformal field theory can be applied. Therefore we can use the ratio of left hand side of (\ref{energy3-4}) to identify the phase boundary.
 Let us call the left-hand side of eq.(\ref{energy3-4}) $Rt$,
 \begin{equation}\label{ratio}
 Rt \equiv  \left( \left.\cfrac{\cfrac{E_{g}(L+2)}{L+2}-\cfrac{E_{g}(L)}{L}}{\cfrac{E_{g}(L)}{L}-\cfrac{E_{g}(L-2)}{L-2}} \right)
\right/ \left( \cfrac{\cfrac{1}{(L+2)^{2}}-\cfrac{1}{L^{2}}}{\cfrac{1}{L^{2}}-\cfrac{1}{(L-2)^{2}}} \right).
 \end{equation}
\par
 In the following, we will examine the dependence of the numerically calculated value of $Rt$ on $D$.

\subsubsection{D-peak shift by $\Delta$}
 Next, we describe how to estimate the phase boundaries by obtaining the dependence of the value of $Rt$ (\ref{ratio}) on $D$ for each $\Delta$.
 
\begin{figure}[hbtp]
 \centering
 \includegraphics[keepaspectratio, scale=0.65]
      {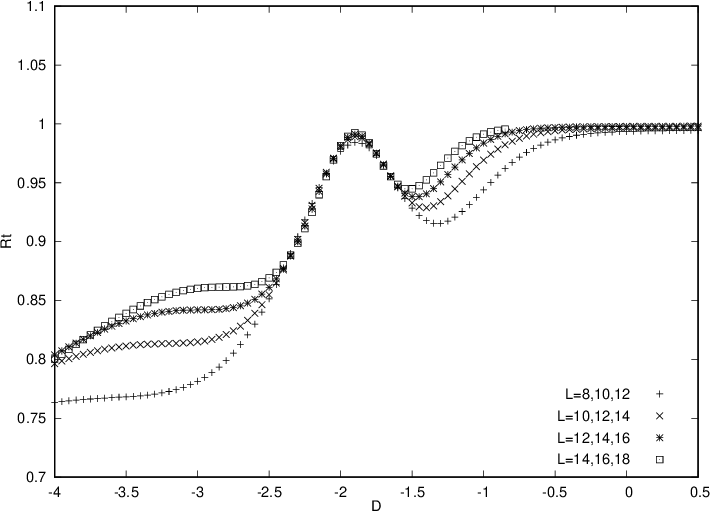}
 \caption{$Rt$ for each different size $L$ are plotted, at $\Delta=0.1$. $Rt$ is defined by eq (21).}
\label{0.1}
\end{figure}

\begin{figure}[hbtp]
 \centering
 \includegraphics[keepaspectratio, scale=0.65]
      {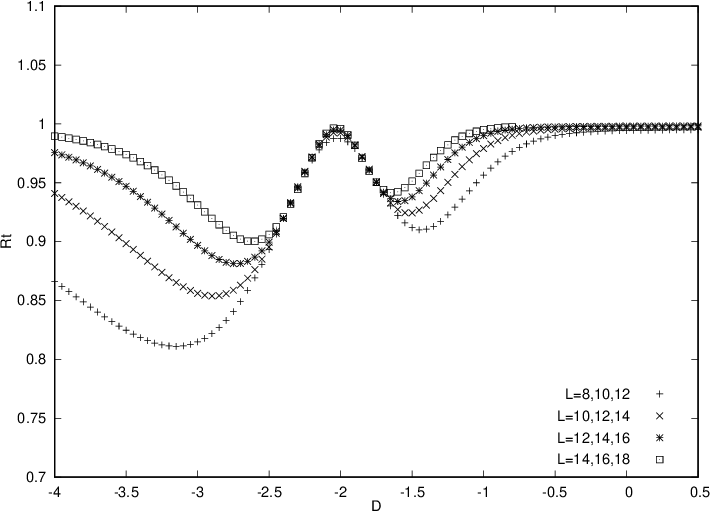}
 \caption{$\Delta=0$}
\label{0}
\end{figure}

\begin{figure}[hbtp]
 \centering
 \includegraphics[keepaspectratio, scale=0.65]
      {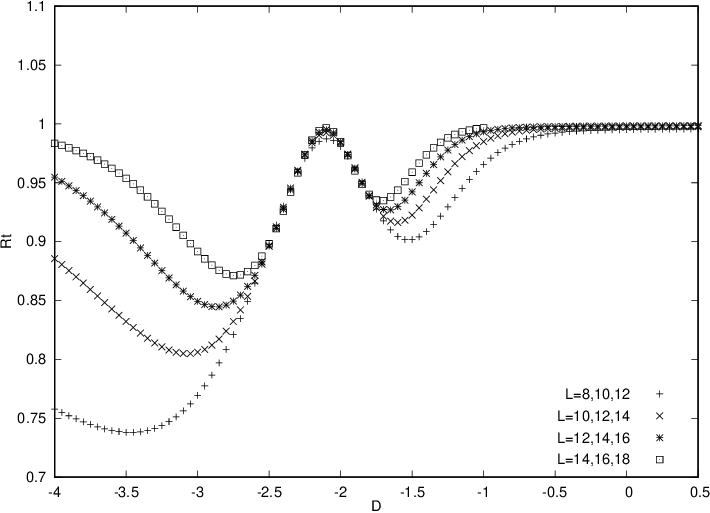}
 \caption{$\Delta=-0.1$}
\label{-0.1}
\end{figure}
FIG. \ref{0.1} shows the $D$ dependence of $Rt$ at $\Delta =0.1$. It can be read that the value of $Rt$ is close to 1 at $D \sim -2$. This corresponds to the N\'eel-Haldane boundary, which can be interpreted as a CFT with central charge $c =1/2$. At this time, the correlation length is expected to behave,

\begin{equation}\label{hn}
\xi_{z}\propto |D-D_{HN}(\Delta)|^{-1},
\end{equation}
where $\xi_{z}$ was defined in Sec.2, and $D_{HN}(\Delta)$ is a Haldane-N\'eel phase boundary.\par

Next we show the $D$ dependence of $Rt$ in Fig.\ref{0} ($\Delta=0$) and Fig. \ref{-0.1} ($\Delta=-0.1$), respectively. Similarly, the value of $Rt$ peaks around $D\sim -2$. From the 2D Ising universality class, at this time, the correlation lengths behave,

\begin{equation}\label{xy1}
\xi_{z}\propto \left(D-D_{XY12}(\Delta)\right)^{-1}, (XY1)
\end{equation}
and
\begin{equation}\label{xy2}
\xi_{1}\propto \left(D_{XY12}(\Delta)-D\right)^{-1}, (XY2)
\end{equation}
where $\xi_{z}$ and $\xi_{1}$  were defined in Sec.2, and $D_{XY12}(\Delta)$ is a phase boundary between XY1 and XY2 phase.\par

 As a further example, data estimated by extrapolation for the value of the peak at $\Delta = 0$ is shown in Fig.8. Here we extrapolated $D$ as $D_{peak}(L) = D_{peak}(\infty) + \beta_{1} \frac{1}{L^{2}}+ \beta_{2} \frac{1}{L^{4}}$ \cite{Cardy-1986},\cite{Reinicke-1986}. We extrapolate the tetracritical point as $ \Delta = 0$, $D = -2.03343$.\par

\begin{figure}[hbtp]
 \centering
 \includegraphics[keepaspectratio, scale=0.65]
      {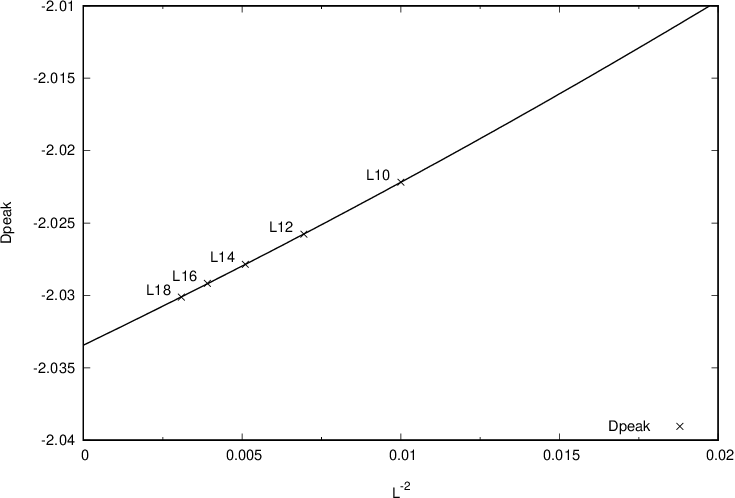}
 \caption{The extrapolation of the $D$'s peak  at $\Delta=0$ for each set of different sizes ($L = 8,10,12$),($L = 10,12,14$),($L = 12,14,16$),($L = 14,16,18$),($L = 16,18,20$).}
\label{peak}
\end{figure}

In summary, from the analysis such as FIG. \ref{0.1}-\ref{-0.1} and FIG. \ref{peak}, the Haldane-N\'eel phase boundary and the XY1-XY2 phase boundary are determined commonly with the peak of $Rt$.\par

The Haldane-large $D$ phase boundary belongs to the Gaussian fixed line with central charge $c=1$.
The Gaussian fixed line associated with BKT lines form a multicritical point,
and renormalization flows become very slow or extremely large $\xi$ in the neighborhood of this BKT multicritical point, which explains the peculiar behavior of $Rt$ in the Haldane phase close to  XY1 and large $D$ phases in Fig.\ref{0.1}.  
Although it is difficult to determine the BKT transition lines and the Gaussian fixed line with the present method, it is valid to apply
the level spectroscopy \cite{Kitazawa-Hijii-Nomura-2003},\cite{Nomura-1995} for the BKT transitions, 
and the Kitazawa's twisted boundary method \cite{Kitazawa-1997}  for the Gaussian fixed line.
By using these methods, Haldane, large $D$ and XY1 phase boundaries were obtained by Chen et. al.\cite{Chen-Hida-Sanctuary-2003}. Thus in this paper we concentrate on the XY1-XY2 and Haldane-N\'eel phase boundaries.

\subsection{Mixed case of finite and infinite correlation lengths}

Next let us consider the mixed case that one correlation length $\xi$ is finite, and another correlation length is infinite. The relationship assumed in this case is

\begin{equation}\label{energy4-1}
\frac{E_{g}(L)}{L}=\epsilon_{\infty}-\frac{\pi v c}{6L^{2}}-k e^{-L/\xi},
\end{equation}
where $k$ is a proportionality constant.\par
 Using different sizes $L$,
\begin{equation}\label{energy4-2}
\begin{split}
\frac{E_{g}(L+2)}{L+2}-\frac{E_{g}(L)}{L}
=
&-\frac{\pi}{6} ( \frac{1}{(L+2)^{2}}-\frac{1}{L^{2}} ) v c \\
&- k ( e^{-(L+2)/\xi}-e^{-L/\xi}) \\
=
&-\frac{\pi}{6} ( \frac{1}{(L+2)^{2}}-\frac{1}{L^{2}} ) v c \\
&- k e^{-L/\xi}( e^{-2/\xi}-1).
\end{split}
\end{equation}

and when $L \gg  \xi$ or $L \sim  \xi$, above relation (\ref{energy4-2}) is 
 
 \begin{equation}\label{energy4-3}
 \frac{E_{g}(L+2)}{L+2}-\frac{E_{g}(L)}{L}
\sim -\frac{\pi}{6} \left( \frac{1}{(L+2)^{2}}-\frac{1}{L^{2}} \right) v c
\end{equation}
\par
\begin{figure}[hbtp]
 \centering
 \includegraphics[keepaspectratio, scale=0.60]
      {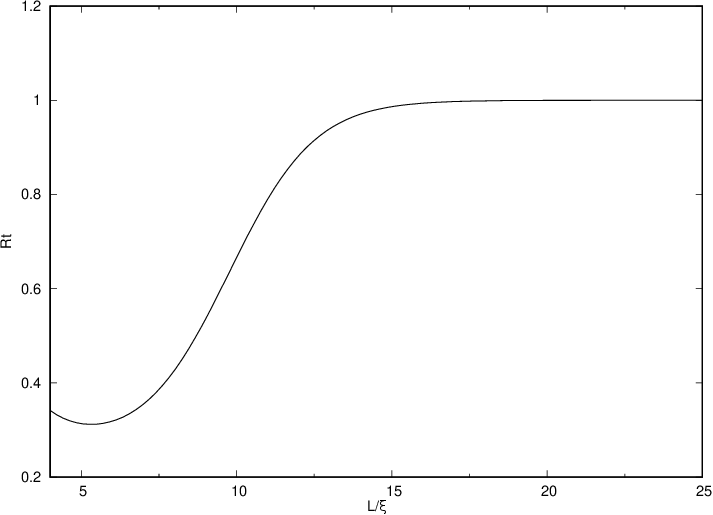}
 \caption{example by eq.(\ref{energy4-2}) : $\xi =2.0, v = 1.0 ,k =1.0, c=1 $ }
\label{example}
\end{figure}

 Again considering $Rt$, for example, when $\xi = 2.0, v = 1.0, k = 1.0$, and $c = 1$, taking $L/\xi$ on the horizontal axis, we obtain a relationship as shown in FIG. \ref{example}. This is consistent with the results of Sec.III-B.

\subsection{Massive case}
In the off-critical regions such the Haldane and N\'eel phases, all the correlation lengths are finite, thus the previous considerations do not hold. In this case, it is expected that the ground state energy behaves as
\begin{equation}\label{energy5}
\frac{E_{g}(L)}{L}={\epsilon}_{\infty}-Ae^{-L/{\xi}}.
\end{equation}
where $A$ is a coefficient and $\xi$ is the longest one of the correlation lengths. In this case, $Rt \approx \exp(-2/\xi)$, thus off-critical behavior in FIG. \ref{0.1} can be explained. This is also consistent with the results of Sec.III-B.

\section{Discussions}
In this section, we discuss symmetry aspects near the tetracritical point. And we compare previous research works with the present one.

\subsection{Reinterpretation of Schulz's discussion}
Reflecting on Schulz's work\cite{Schulz-1986}, he first rewrote the S=1 XXZ model (\ref{hamiltonian}) as the S=1/2 spin ladder system with ferromagnetic rung couplings.\par
\begin{figure}[hbtp]
 \centering
 \includegraphics[keepaspectratio, scale=0.46]
      {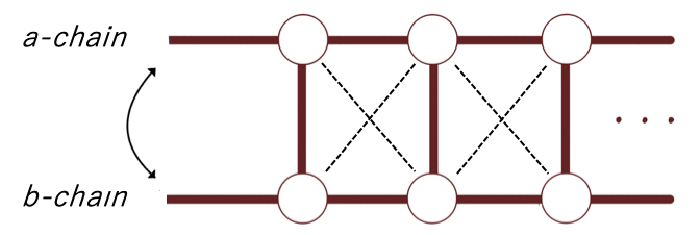}
 \caption{The rung-inversion symmetry}
\label{ladder}
\end{figure}

Then, by using the bosonization method, he derived two independent effective field models.
We consider that the important point is the rung-inversion symmetry in the spin ladder system as shown in FIG. \ref{ladder},
which survives after approximations.\par
Thus eigenstates are divided into a rung-inversion symmetric sector and an antisymmetric sector. With this symmetry the four phases can intersect at one-point. Note that this rung-inversion symmetry is also important in the proof of Kitazawa et al\cite{Kitazawa-Hijii-Nomura-2003}.

\par

\subsection{Gibbs's phase rule}
In general, for two intensive variables,
it is known that the Gibbs's phase rule holds
\begin{equation}
F=C-P+2,
\end{equation}
where $F$ is the number of degrees freedom,$C$ is the number of components, and $P$ is the number of phases.\par
In our cases, there exist two independent components as explained in Sec. IV-A,
thus for four phases we obtain $F=0$, which means a point.

\subsection{Symmetry and universality}
We discuss the difference between the symmetry of $D=0$ in 
(\ref{eq:BLBQ-w-sigle-ion-anisotropy})
and that of $\Delta=0$ in
(\ref{hamiltonian}).
On the line $D=0$ in
(\ref{eq:BLBQ-w-sigle-ion-anisotropy}), the model
has an SO(3) symmetry, thus the universality at the TB point is explained as 
      the ${\rm SO(3)_{1}}$ or the ${\rm SU(2)_{2}}$
      Wess-Zumino-Witten (WZW) model \cite{Affleck-Haldane-1987}
      with central charge $c=3/2$.
      In the ${\rm SO(3)_{1}}$ WZW model, there are several relevant fields,
      the important one of them has scaling dimension $x=1$
with SO(3) and translational invariance.
      Relaxing the symmetry to SO(2), there exists another scaling field,
      and these two fields $x=1$ explain the tetracritical behavior near the TB point.

Returning to the model (\ref{hamiltonian}), on the line $\Delta=0$ 
it has the special SU(2) symmetry as shown in (\ref{s2}) and (\ref{s4}). 
Nonetheless it should be noted that the special SU(2) symmetry in the magnetization $m$ even sector and that
in the $m$ odd sector
are independent \cite{Kitazawa-Hijii-Nomura-2003}
(more precisely, this symmetry depends on boundary conditions; for open boundary condition, the special SU(2) symmetry simply holds; for periodic boundary condition,
one should extend twisted boundary conditions.). 
The special SU(2) symmetry in the even $m$ sector induces the ${\rm SU(2)_{1}}$
WZW model with $c=1$, or the BKT phase boundary.
About the special SU(2) symmetry in the odd $m$ sector,
it may be related with a different phase transition at the tetracritical point.

\subsection{Previous numerical methods for XY1-XY2 boundary.}
At the Haldane-N\'eel phase boundary, the phenomenological renormalization group (PRG)\cite{Fisher and Barber} is valid because this is a second-order phase transition of order-disorder, which is a well established method\cite{Chen-Hida-Sanctuary-2003}.\par

For the XY1-XY2 phase boundary, the following magnetization-based methods have been used. The difference between the excitation energy and the ground energy is denoted by $\Delta E(m) =E(m) -E_{g}$. 

Then we can define,
\begin{equation}
mrt \equiv \frac{{\Delta}E(m=2)}{{\Delta}E(m=1)}
\end{equation}
 
 \par
 
From Schulz's discussion \cite{Schulz-1986} and CFT, one can derive that $mrt =4$ in XY1 phase and $mrt = 0$ in XY2 phase. Thus, numerically in \cite{Chen-Hida-Sanctuary-2003} they used $mrt = 1$ to determine XY1-XY2 phases, whereas in \cite{Tonegawa-Okamoto-Nomura-Sakai-2022} , they used $mrt =2$.\par

One cannot say whether method \cite{Chen-Hida-Sanctuary-2003} or \cite{Tonegawa-Okamoto-Nomura-Sakai-2022} is more suitable to determine the phase boundary. In fact, both methods can not be justified directly from Schulz's study \cite{Schulz-1986}.\par

Note that the jump of $mrt$ may be used to detect XY1-XY2 phase boundary, 
this may not be related with the Haldane-N\'eel phase boundary.

 \subsection{Different CFT velocities}

   In eq.(\ref{energy3-1}) of  Sec. III-B, we have discussed that there may be two different CFT velocities.
   Such phenomena have been known in the spin-charge separation of the 1D Hubbard model\cite{Woynarovich-1989,Frahm-Korepin-1990},
   which is Bethe-Ansatz solvable.
   In the case of less than half filling,
   both charge and spin excitations are gapless,
   and they are described as two $c=1$ CFT with different velocities $v_{c}, v_{s}$.   

As for the spin case, multiple CFT velocities have been reported
  in the S=7/2 integrable XXZ spin chain \cite{Frahm-Yu-1990}, 
  and in the chiral spin liquid of the S=1/2 spin ladder under magnetic field \cite{Frahm-Rodenbeck-1997}.

\section{Conclusion}
In this paper, we have studied the S=1 XXZ spin chain with single ion anisotropy,
    and proposed a unified numerical method to determine
    the XY1-XY2 phase boundary and the  Haldane-N\'eel phase boundary.
    These two phase boundaries continuously connect,
    thus combining with the BKT line already known on the $\Delta=0$,
    there is a tetracritical point as expected in \cite{Schulz-1986}.
    In addition, the XY1-XY2 phase boundary and the Haldane-N\'eel phase boundary are smoothly connected(i.e., differential is also continuous).\par
        On the universality class of the tetracritical point,
    from the effective central charge, we discussed that this consists of the two independent field theories with different spin wave velocities.
    They can be explained with the rung inversion symmetry of the two leg spin ladder.\par
    Finally, it is known that the Bose-Hubbard model,
which is used to describe ultracold bosons in optical lattices,
can be mapped onto the S=1 XXZ chain with single ion anisotropy
\cite{
Altman-Auerbah-2002,
  Lee-Lee-Yang-2007,Berg-Torre-Giamarchi-Altman-2008,Cuzzuol-Barbiero-Montorsi-2024}.
In the work \cite{Cuzzuol-Barbiero-Montorsi-2024}, they imposed three body constraint $(b_{R}^{\dagger})^{3}=0$, thus the on-site attractive interaction becomes possible.
Therefore our present study may contribute to understand the Bose-Hubbard model.

\begin{acknowledgments}
We are grateful to K. Okamoto for discussions on this work.
About the spin nematic phase, K. N. are taught from T. Momoi.
We thank H. Katsura for references about the Takhtajan-Babujian point,
and those about multiple CFT velocities in spin models.
We also thank I. Maruyama for useful comments.
  K. N. is supported by Japan Society for the promotion of Science KAKENHI, GRANT No. 21H05021. S.S would like to acknowledge the support of the Kyushu University Leading Human Resources Development Fellowship Program. The numerical diagonalization program used in this study is based on the program developed by S.Moriya from TiTpack and Kobe Pack.

\end{acknowledgments}

\appendix

\section{Perturbation theory and effective hamiltonian}
Using ladder operators $S_{i}^{\pm}=S_{i}^{x} \pm i S_{i}^{y}$, Hamiltonian (\ref{hamiltonian}) is shown as
\begin{equation}
H=\sum_{j}(\frac{1}{2}(S_{j}^{+} S_{j+1}^{-}+S_{j}^{-} S_{j+1}^{+})+{\Delta}{S}_{j}^{z}{S}_{j+1}^{z}+D({S}_{j}^{z})^{2}).
\end{equation}
Denote the basis of the state space of the $j$th particle as

\begin{equation}
|\uparrow>=\left(\begin{array}{l}
1 \\
0 \\
0
\end{array}\right),\  \ \
|0>=\left(\begin{array}{l}
0 \\
1 \\
0
\end{array}\right),\ \ \ \
|\downarrow>=\left(\begin{array}{l}
0 \\
0 \\
1
\end{array}\right),
\end{equation}
so that, the representation matrix of the spin operators are
\begin{equation}
\begin{split}
S^{x}&=\frac{1}{\sqrt{2}}\begin{pmatrix} 0 & 1 & 0\\ 1 & 0 & 1\\ 0 & 1 & 0\end{pmatrix}\ ,\ \ S^{y}=\frac{1}{\sqrt{2}}\begin{pmatrix} 0 & -i & 0\\ i & 0 & -i\\ 0 & i & 0\end{pmatrix}\ ,\ \\
S^{z}&=\begin{pmatrix} 1 & 0 & 0\\ 0 & 0 & 0\\ 0 & 0 & -1\end{pmatrix}.
\end{split}
\end{equation}

\begin{equation}
S^{\pm}=S^{x} \pm iS^{y}
\end{equation}

For $D \rightarrow  - \infty$, the terms not including $D$ in Hamiltonian (A1) can be treated as a perturbation. Thus, we denote\par

\begin{equation}
H = D({S}_{j}^{z})^{2} + \lambda V,
\end{equation}
and the term containing $V$ is considered a perturbation.

The eigenvalues of $D({S}_{j}^{z})^{2}$ for the $j$ th site are 0 or $D$. When considering $D \rightarrow \infty$, $D$ corresponds to the ground state energy.The eigenvalue $D$ is doubly degenerate, with states $|\uparrow>$ and $|\downarrow>$. The entire system will have $2^{N}$ degeneracies for one eigenvalue $ND$.\par
When the $j$ th $j+1$ th states are $|\uparrow>$ and $|\downarrow>$, respectively, and the state of the entire system is denoted as $|\uparrow_{j}\downarrow_{j+1}>=|\uparrow_{j}> \otimes |\downarrow_{j+1}>$, then 
\begin{equation}
|\uparrow_{j}\uparrow_{j+1}>\ \ ,
|\uparrow_{j}\downarrow_{j+1}>\ \ ,
|\downarrow_{j}\uparrow_{j+1}>\ \ ,
|\downarrow_{j}\downarrow_{j+1}>
\end{equation}
have all the same eigenvalues.Here, 
\begin{equation}
V=\sum_{j}(J_{i}({S}_{j}^{x}{S}_{j+1}^{x}+{S}_{j}^{y}{S}_{j+1}^{y})+{\Delta}{S}_{j}^{z}{S}_{j+1}^{z})
\end{equation}

 is represented as $V=\sum_{j}V_{j}$, where $V_{j}=J_{i}({S}_{j}^{x}{S}_{j+1}^{x}+{S}_{j}^{y}{S}_{j+1}^{y})+{\Delta}{S}_{j}^{z}{S}_{j+1}^{z})$. Thus, we calculate $<\uparrow_{j}\uparrow_{j+1}| v |\uparrow_{j}\uparrow_{j+1}>$ to obtain the matrix representation of $v$. However, since we are considering only the ground state, we can calculate the components of  $ 2\times 2$ matrix as follows.
 For example,

\begin{equation}
\begin{split}
{S}_{j}^{z}{S}_{j+1}^{z} |\uparrow_{j}\uparrow_{j+1} \rangle &=
\begin{pmatrix} 1 & 0 & 0\\ 0 & 0 & 0\\ 0 & 0 & -1\end{pmatrix} 
\left(\begin{array}{l}
1 \\
0 \\
0
\end{array}\right)
\otimes
\begin{pmatrix} 1 & 0 & 0\\ 0 & 0 & 0\\ 0 & 0 & -1\end{pmatrix} 
\left(\begin{array}{l}
1 \\
0 \\
0
\end{array}\right) \\
&=
\left(\begin{array}{l}
1 \\
0 \\
0
\end{array}\right)
\otimes
\left(\begin{array}{l}
1 \\
0 \\
0
\end{array}\right) = |\uparrow_{j}\uparrow_{j+1} \rangle
\end{split}
\end{equation}
thus,
\begin{equation}
<\uparrow_{j}\uparrow_{j+1}| \Delta {S}_{j}^{z}{S}_{j+1}^{z}  |\uparrow_{j}\uparrow_{j+1}>
= \Delta 
\end{equation}
Calculating the other components in the same way, the matrix representation of the Hamiltonian by perturbation of the first order $H^{(1)}_{eff}$is
\begin{equation}
 H^{(1)}_{eff}=\Delta \ \ 
\begin{blockarray}{ccccc}
\ & \ & \ & \ & \ \\
\begin{block}{(cccc)c}
  1 & 0 & 0 & 0 & \ |\uparrow \uparrow >\\
  0 & -1 & 0 & 0 & \ |\uparrow \downarrow >\\
  0 & 0 & -1 & 0 & \ |\downarrow \uparrow >\\
  0 & 0 & 0 & 1 & \ |\downarrow \downarrow >\\
\end{block}
\end{blockarray}
\end{equation}

On the other hand, when $S=1/2$, the matrix representation of the spin operator is
\begin{equation}
T^{x}=\frac{1}{2}\begin{pmatrix} 0 & 1 \\ 1 & 0 \end{pmatrix}\ ,\ \
T^{y}=\frac{1}{2}\begin{pmatrix} 0 & -i \\i & 0 \end{pmatrix}\ ,\ \
T^{z}=\frac{1}{2}\begin{pmatrix} 1 & 0 \\ 0 & -1\end{pmatrix}.\ \ \
\end{equation}
Thus,after simple calculations, we obtain
\begin{equation}
 T^{z}_{j}T^{z}_{j+1}=\frac{1}{4} \ \ 
\begin{blockarray}{ccccc}
\ & \ & \ & \ & \ \\
\begin{block}{(cccc)c}
  1 & 0 & 0 & 0 & \ |\uparrow \uparrow >\\
  0 & -1 & 0 & 0 & \ |\uparrow \downarrow >\\
  0 & 0 & -1 & 0 & \ |\downarrow \uparrow >\\
  0 & 0 & 0 & 1 & \ |\downarrow \downarrow >\\
\end{block}
\end{blockarray}
\end{equation}

If we map the S=1 state space to the S=1/2 state space, we can express the effective Hamiltonian as 
\begin{equation}
H^{(1)}_{eff}=4\Delta \sum_{j}T^{z}_{j}T^{z}_{j+1},
\end{equation}
 using the expression S=1/2.
Furthermore, let us consider second-order perturbations. Since the states that can transition are limited if we consider the selection rule, we only need to consider the case where the intermediate state is $|00 \rangle$.\par
Calculating for each term,
 \begin{equation}
 \frac{<\uparrow_{j}\downarrow_{j+1}| \frac{1}{2} {S}_{j}^{+}{S}_{j+1}^{-}  |00>
 <00| \frac{1}{2} {S}_{j}^{+}{S}_{j+1}^{-}  |\downarrow_{j}\uparrow_{j+1}>}{-2D}
 =\frac{1}{-2D}
 \end{equation}
 
 \begin{equation}
 \frac{<\uparrow_{j}\downarrow_{j+1}| \frac{1}{2} {S}_{j}^{+}{S}_{j+1}^{-}  |00>
 <00| \frac{1}{2} {S}_{j}^{+}{S}_{j+1}^{-}  |\uparrow_{j}\downarrow_{j+1}>}{-2D}
 =\frac{1}{-2D}
 \end{equation}
 
 \begin{equation}
 \frac{<\downarrow_{j}\uparrow_{j+1}| \frac{1}{2} {S}_{j}^{+}{S}_{j+1}^{-}  |00>
 <00| \frac{1}{2} {S}_{j}^{+}{S}_{j+1}^{-}  |\uparrow_{j}\downarrow_{j+1}>}{-2D}
 =\frac{1}{-2D}
 \end{equation}
 
 \begin{equation}
 \frac{<\downarrow_{j}\uparrow_{j+1}| \frac{1}{2} {S}_{j}^{+}{S}_{j+1}^{-}  |00>
 <00| \frac{1}{2} {S}_{j}^{+}{S}_{j+1}^{-}  |\downarrow_{j}\uparrow_{j+1}>}{-2D}
 =\frac{1}{-2D}
 \end{equation}
 we find that all these others are zero, therefore the Hamiltonian of the second-order perturbation $H^{(2)}_{eff}$ is 
 \begin{equation}
 H^{(2)}_{eff}=\frac{1}{2|D|} \ \ 
\begin{blockarray}{ccccc}
\ & \ & \ & \ & \ \\
\begin{block}{(cccc)c}
  0 & 0 & 0 & 0 & \ |\uparrow \uparrow >\\
  0 & -1 & -1 & 0 & \ |\uparrow \downarrow >\\
  0 & -1 & -1 & 0 & \ |\downarrow \uparrow >\\
  0 & 0 & 0 & 0 & \ |\downarrow \downarrow >\\
\end{block}
\end{blockarray}
\end{equation}

Further transforming, we obtain
\begin{equation}
H^{(2)}_{eff}=-\frac{1}{4|D|}
\begin{pmatrix} 1 & 0 & 0 & 0\\ 0 & 1 & 0 & 0\\ 0 & 0 & 1 & 0\\ 0 & 0 & 0 & 1
\end{pmatrix}
+\frac{1}{4|D|}
\begin{pmatrix} 1 & 0 & 0 & 0\\ 0 & -1 & -2 & 0\\ 0 & -2 & -1 & 0\\ 0 & 0 & 0 & 1
\end{pmatrix}
\end{equation}
where, the 1st term is constant.\par
 Here again, for the $S=1/2$ case, the matrix representation of the XXZ model is shown in 
 \begin{equation}
 \begin{split}
 \frac{1}{2}J(T_{j}^{+} T_{j+1}^{-}+T_{j}^{-} T_{j+1}^{+})
 &+J_{z}{T}_{j}^{z}{T}_{j+1}^{z}\\
 &=
\begin{blockarray}{ccccc}
\ & \ & \ & \ & \ \\
\begin{block}{(cccc)c}
  \frac{1}{4}J_{z} & 0 & 0 & 0 & \ |\uparrow \uparrow >\\
  0 & -\frac{1}{4}J_{z} & \frac{1}{2}J & 0 & \ |\uparrow \downarrow >\\
  0 & \frac{1}{2}J & -\frac{1}{4}J_{z} & 0 & \ |\downarrow \uparrow >\\
  0 & 0 & 0 & \frac{1}{4}J_{z} & \ |\downarrow \downarrow >\\
\end{block}
\end{blockarray}
\end{split}
\end{equation}

Comparing with the second term of equation (A19), we can put 
\begin{equation}
J_{z}=\frac{1}{|D|}, \ \ \ J=-\frac{1}{|D|}
\end{equation}
Thus, in the second-order perturbation,
\begin{equation}
H^{(2)}_{eff}=\sum_{j}(-\frac{1}{|D|}(T^{x}_{j}T^{x}_{j+1}+T^{y}_{j}T^{y}_{j+1})+\frac{1}{|D|}T^{z}_{j}T^{z}_{j+1})
\end{equation}
 From (A13) and (A22), we obtain
 \begin{equation}
 H_{eff}=\sum_{j}(-\frac{1}{|D|}(T^{x}_{j}T^{x}_{j+1}+T^{y}_{j}T^{y}_{j+1})+(\frac{1}{|D|}+4\Delta)T^{z}_{j}T^{z}_{j+1})
\end{equation}
 Furthermore, performing unitary transformations such as
 \begin{equation}
 T^{x,y}_{j}\rightarrow
\begin{cases}
-T^{x,y}_{j} & :j=2i \\
T^{x,y}_{j} & :j=2i+1
\end{cases}
\end{equation}
we obtain the S=1/2 XXZ model

\begin{equation}
 H_{eff}=\sum_{j}(\frac{1}{|D|}(T^{x}_{j}T^{x}_{j+1}+T^{y}_{j}T^{y}_{j+1})+(\frac{1}{|D|}+4\Delta)T^{z}_{j}T^{z}_{j+1})
\end{equation}
\\ \\
 whose properties are well known.\par
 \ \par
 \


\newpage
\begin{thebibliography}{999}

 \bibitem{Schulz-1986}
H. J. Schulz,
Phase diagrams and correlation exponents for quantum spin chains of arbitrary spin quantum number,
\href{https://doi.org/10.1103/PhysRevB.34.6372}{Phys. Rev. B {\bf 34}, 6372 (1986)}.
 \bibitem{Chen-Hida-Sanctuary-2003}
Chen, K. Hida and B. C. Sanctuary,
Ground-state phase diagram of $S = 1$ XXZ chains with uniaxial single-ion-type anisotropy,
\href{https://doi.org/10.1103/PhysRevB.67.104401}{ Phys Rev. B {\bf 67},  104401 (2003)}.
 \bibitem{Kitazawa-Hijii-Nomura-2003}
A. Kitazawa, K. Hijii, and K. Nomura,
An SU(2) symmetry of the one-dimensional spin-1 XY model,
\href{https://doi.org/10.1088/0305-4470/36/23/104}{J. Phys. A {\bf 36}, L351,(2003)}.
 \bibitem{Nomura-1995}
 K. Nomura,
 Correlation functions of the 2D sine-Gordon model,
 \href{https://doi.org/10.1088/0305-4470/28/19/003}{J. Phys. A {\bf 28},  5451, (1995)}
 \bibitem{Nomura-Kitazawa-1998}
 K. Nomura and A. Kitazawa,
  $SU(2)/Z_{2}$ symmetry of the BKT transition and twisted boundary condition,
  \href{https://doi.org/10.1088/0305-4470/31/36/008}{J. Phys. A {\bf 31},  7341, (1998)}
\bibitem{Kitazawa-1997}
A. Kitazawa,
Twisted boundary conditions of quantum spin chains near the Gaussian fixed points,
\href{https://doi.org/10.1088/0305-4470/30/9/005}{J. Phys. A, {\bf 30}, L285, (1997)}.
\bibitem{Solyom-Ziman-1984}
  J. Solyom and T. Ziman,
  Ground-state properties of axially anisotropic quantum Heisenberg chains,
  \href{https://doi.org/10.1103/PhysRevB.30.3980}{Phys. Rev. B {\bf 30}, 3980 (1984)}.
\bibitem{Schulz-Ziman-1986}
  H. J. Schulz and T. Ziman,
  Finite-length calculations of $\eta$ and phase diagrams of quantum spin chains,
   \href{https://doi.org/10.1103/PhysRevB.33.6545}{Phys. Rev, B {\bf 33}, 6545 (1986)}.
\bibitem{Fisher and Barber}
  M. E. Fisher and M. N. Barber,
  Scaling Theory for Finite-Size Effects in the Critical Region,
\href{https://doi.org/10.1103/PhysRevLett.28.1516}{Phys. Rev. Lett. {\bf 28}, 1516 (1972)}.
\bibitem{Tonegawa-Okamoto-Nomura-Sakai-2022}
T.Tonegawa,K.Okamoto,K.Nomura,and T.Sakai,
Nematic Tomonaga-Luttinger Liquid Phase in an S=1/2 Ferromagnetic-Antiferromagnetic Bond-Alternating Chain,
\href{https://doi.org/10.7566/JPSCP.38.011154}{JPS Conf. Proc. {\bf 38}, 011154, 6 (2023)}.
         
\bibitem{Takhtajan-1982}
  L. A. Takhtajan,
  The picture of low-lying excitations in the isotropic Heisenberg chain of arbitrary spins,
  \href{https://doi.org/10.1016/0375-9601(82)90764-2}{Phys. Lett. A {\bf 87}, 479 (1982)}.
\bibitem{Babujian-1982}
  H. M. Babujian,
  Exact solution of the one-dimensional isotropic Heisenberg chain with arbitrary spins S,
  \href{https://doi.org/10.1016/0375-9601(82)90403-0}{Phys. Lett. A {\bf 90},479 (1982)}.
\bibitem{Babujian-1983}
  H. M. Babujian,
  Exact solution of the isotropic Heisenberg chain with arbitrary spins: Thermodynamics of the model,
  \href{https://doi.org/10.1016/0550-3213(83)90668-5}{Nucl. Phys. B {\bf 215}, 317 (1983)}.

 
 \bibitem{Affleck-Haldane-1987}
  I. Affleck and F. D. M. Haldane,
  Critical theory of quantum spin chains,
  \href{https://doi.org/10.1103/PhysRevB.36.5291}{Phys. Rev. B {\bf 36}, 5291 (1987)}.
\bibitem{Tsvelick-1990}
  A. M. Tsvelik,
  Field-theory treatment of the Heisenberg spin-1 chain,
  \href{https://doi.org/10.1103/PhysRevB.42.10499}{Phys. Rev. B {\bf 42}, 10499 (1990)}.
\bibitem{DeChiara-Lewenstein-Sanpera-2011}
  G. De Chiara, M. Lewenstein, and A. Sanpera,
  Bilinear-biquadratic spin-1 chain undergoing quadratic Zeeman effect,
  \href{https://doi.org/10.1103/PhysRevB.84.054451}{Phys. Rev. B {\bf 84}, 054451 (2011)}.
\bibitem{Lepri-DeChiara-Sanpera-2013}
  L. Lepori, G. De Chiara, and A. Sanpera,
  Scaling of the entanglement spectrum near quantum phase transitions
  \href{https://doi.org/10.1103/PhysRevB.87.235107}{Phys. Rev. B {\bf 87}, 235107 (2013)}.

\bibitem{Ginsparg}
         P. H. Ginsparg, Applied Conformal Field Theory,
         \href{https://doi.org/10.48550/arXiv.hep-th/9108028}{arXiv:hep-th/9108028}.
\bibitem{Kitazawa-Nomura-1997}
 A. Kitazawa and K. Nomura,
 Critical Properties of S=1 Bond-Alternating XXZ Chains and Hidden Z 2× Z 2 Symmetry,
\href{https://doi.org/10.1143/JPSJ.66.3944}{JPSJ. A {\bf 66},  p.3944, (1997)}.
\bibitem{Blote-1986}
H. W. J. Bl\"{o}te, J. L. Cardy and M. P. Nightingale,
Conformal invariance, the central charge, and universal finite-size amplitudes at criticality,
\href{https://doi.org/10.1103/PhysRevLett.56.742}{Phys.Rev. Lett. {\bf 56} 742 ,(1986)}.
\bibitem{Affleck-1986}
I. Affleck,
Universal term in the free energy at a critical point and the conformal anomaly,
\href{https://doi.org/10.1103/PhysRevLett.56.746}{Phys. Rev. Lett. {\bf 56}, 746 (1986)}.
\bibitem{Cardy-1986}
J. L. Cardy,
Operator content of two-dimensional conformally invariant theories,
\href{https://doi.org/10.1016/0550-3213(86)90552-3}{Nucl. Phys. B {\bf 270} [FS16], 186 (1986)}.
\bibitem{Reinicke-1986}
P. Reinicke,
Analytical and non-analytical corrections to finite-size scaling,
\href{https://doi.org/10.1088/0305-4470/20/15/044}{J. Phys. A: Math. Gen. {\bf 20}, 5325 (1987)}.
\bibitem{Woynarovich-1989}
  F. Woynarovich,
  Finite-size effects in a non-half-filled Hubbard chain,
  \href{https://doi.org/10.1088/0305-4470/22/19/017}{J. Phys. A {\bf 22}, 4243 (1989)}.
\bibitem{Frahm-Korepin-1990}
  H. Frahm and V. E. Korepin,
  Critical exponents for the one-dimensional Hubbard model,
  \href{https://doi.org/10.1103/PhysRevB.42.10553}{Phys. Rev. B {\bf 42}, 10553 (1990)}. 

  \bibitem{Frahm-Yu-1990}
  H. Frahm and Nai-C Yu,
  Finite-size effects in the integrable XXZ Heisenberg model with arbitrary spin,
 \href{https://doi.org/10.1088/0305-4470/23/11/032}{J. Phys. A: Math. Gen. {\bf 23}, 2115 (1990)}. 
\bibitem{Frahm-Rodenbeck-1997}
  H. Frahm and C. R\"{o}denbeck,
  Properties of the chiral spin liquid state in generalized spin ladders,
 \href{https://doi.org/10.1088/0305-4470/30/13/005}{J. Phys. A: Math. Gen. {\bf 30}, 4467 (1997)}. 
\bibitem{Altman-Auerbah-2002}
  E. Altman and A. Auerbach,
  Oscillating Superfluidity of Bosons in Optical Lattices,
  \href{https://doi.org/10.1103/PhysRevLett.89.250404}{Phys. Rev. Lett. {\bf 89}. 250404 (2002)}.
\bibitem{Lee-Lee-Yang-2007}
  Yu-Wen Lee, Yu-Li Lee, and Min-Fong Yang,
  Low-energy effective theory for one-dimensional lattice bosons near integer filling,
  \href{https://doi.org/10.1103/PhysRevB.76.075117}{Phys. Rev. B {\bf 76}, 075117 (2007)}. 
\bibitem{Berg-Torre-Giamarchi-Altman-2008}
  E. Berg, E. G. Dalla Torre, T. Giamarchi, and E. Altman,
  Rise and fall of hidden string order of lattice bosons,
  \href{https://doi.org/10.1103/PhysRevB.77.245119}{Phys. Rev. B {\bf 77}, 245119 (2008)}. 
\bibitem{Cuzzuol-Barbiero-Montorsi-2024}
  N. Cuzzuol, L. Barbiero, and A. Montorsi,
  Nonlocal order parameter of pair superfluids,
 \href{https://doi.org/10.48550/arXiv.2404.15972}{arXiv:2404.15972}.
       \end{thebibliography}
\end{document}